\documentclass[3p,twocolumn]{elsarticle}
\usepackage{amsmath, amssymb, amsfonts, amsthm}
\usepackage{graphicx}
\usepackage{gastex}
\usepackage{wrapfig}
\theoremstyle{plain}
\usepackage{hyperref}
\usepackage{lineno}

\newtheorem*{theorem}{Theorem}
\newtheorem{lemma}{Lemma}
\newtheorem*{conjecture}{Conjecture}
\theoremstyle{definition}
\newtheorem*{example}{Example}
\newcommand{\LCE}{\mathit{LCE}}
\newcommand{\lcp}{{\sf lcp}}
\newcommand{\sa}{{\sf sa}}
\newcommand{\nodes}{{\sf nds}}
\newcommand{\nil}{\mathbf{nil}}

\journal{Information Processing Letters}
\begin{document}

\begin{frontmatter}

\title{Computing Runs on a General Alphabet}

\author{Dmitry Kosolobov} 
\address{Ural Federal University, Ekaterinburg, Russia}

\begin{abstract}
We describe a RAM algorithm computing all runs (maximal repetitions) of a given string of length $n$ over a general ordered alphabet in $O(n\log^{\frac{2}3} n)$ time and linear space. Our algorithm outperforms all known solutions working in $\Theta(n\log\sigma)$ time provided $\sigma = n^{\Omega(1)}$, where $\sigma$ is the alphabet size. We conjecture that there exists a linear time RAM algorithm finding all runs.
\end{abstract}


\end{frontmatter}


\section{Introduction}

Repetitions in strings are fundamental objects in both stringology and combinatorics on words. In some sense the notion of \emph{run}, introduced by Main~\cite{Main}, allows to grasp the whole repetitive structure of a given string in a relatively simple form. Informally, a run of a string is a maximal periodic substring that is at least as long as twice its minimal period (the precise definition follows). In~\cite{KolpakovKucherov} Kolpakov and Kucherov showed that any string of length $n$ contains $O(n)$ runs and proposed an algorithm computing all runs in linear time on an integer alphabet $\{0,1,\ldots, n^{O(1)}\}$ and $O(n\log\sigma)$ time on a general ordered alphabet, where $\sigma$ is the number of distinct letters in the input string. Recently, Bannai et al. described another interesting algorithm computing all runs in $O(n\log\sigma)$ time~\cite{BannaiIInenagaNakashimaTakedaTsuruta}. Modifying the approach of \cite{BannaiIInenagaNakashimaTakedaTsuruta}, we prove the following theorem.
\begin{theorem}
For a general ordered alphabet, there is an algorithm that computes all runs in a string of length $n$ in $O(n\log^{\frac{2}{3}} n)$ time and linear space.
\end{theorem}
This is in contrast to the result of Main and Lorentz \cite{MainLorentz} who proved that any algorithm deciding whether a string over a general \emph{unordered} alphabet has at least one run requires $\Omega(n\log n)$ comparisons in the worst case.

Our algorithm outperforms all known solutions when the number of distinct letters in the input string is sufficiently large (e.g., $\sigma = n^{\Omega(1)}$). It should be noted that the algorithm of Kolpakov and Kucherov can hardly be improved in a similar way since it strongly relies on a structure (namely, the Lempel--Ziv decomposition) that cannot be computed in $o(n\log\sigma)$ time on a general ordered alphabet (see \cite{Kosolobov}).

Based on some theoretical observations of \cite{Kosolobov}, we conjecture that one can further improve our result.
\begin{conjecture}
For a general ordered alphabet, there is a linear time algorithm computing all runs.
\end{conjecture}

\section{Preliminaries}

A \emph{string of length $n$} over an alphabet $\Sigma$ is a map $\{1,2,\ldots,n\} \mapsto \Sigma$, where $n$ is referred to as the length of $w$, denoted by $|w|$. We write $w[i]$ for the $i$th letter of $w$ and $w[i..j]$ for $w[i]w[i{+}1]\ldots w[j]$. A string $u$ is a \emph{substring} (or a \emph{factor}) of $w$ if $u=w[i..j]$ for some $i$ and $j$. The pair $(i,j)$ is not necessarily unique; we say that $i$ specifies an \emph{occurrence} of $u$ in $w$. A string can have many occurrences in another string. A substring $w[1..j]$ (respectively, $w[i..n]$) is a \emph{prefix} (respectively, \emph{suffix}) of $w$. An integer $p$ is a \emph{period} of $w$ if $0 < p \le |w|$ and $w[i] = w[i{+}p]$ for all $i=1,\ldots,|w|{-}p$; $p$ is the \emph{minimal period} of $w$ if $p$ is the minimal positive integer that is a period of $w$. For integers $i$ and $j$, the set $\{k\in \mathbb{Z} \colon i \le k \le j\}$ (possibly empty) is denoted by $[i..j]$. Denote $[i..j) = [i..j{-}1]$ and $(i..j] = [i{+}1..j]$.

A \emph{run} of a string $w$ is a substring $w[i..j]$ whose period is at most half of the length of $w[i..j]$ and such that both substrings $w[i{-}1..j]$ and $w[i..j{+}1]$, if defined, have strictly greater minimal periods than $w[i..j]$.

We say that an alphabet is \emph{general} and \emph{ordered} if it is totally ordered and the only allowed operation is comparing two letters. Hereafter, $w$ denotes the input string of length $n$ over a general ordered alphabet.

In the \emph{longest common extension ($\LCE$)} problem one has to preprocess $w$ for queries $\LCE(i,j)$ returning for given positions $i$ and $j$ of $w$ the length of the longest common prefix of the suffixes $w[i..n]$ and $w[j..n]$. It is well known that one can perform the $\LCE$ queries in constant time after preprocessing $w$ in $O(n\log\sigma)$ time, where $\sigma$ is the number of distinct letters in $w$ (e.g., see \cite{HarelTarjan}). It turns out that the time consumed by the $\LCE$ queries is dominating in the algorithm of \cite{BannaiIInenagaNakashimaTakedaTsuruta}; namely, one can prove the following lemma.
\begin{lemma}[{see \cite[Alg. 1 and Sect. 4.2]{BannaiIInenagaNakashimaTakedaTsuruta}}]
Suppose we can answer in an online fashion any sequence of $O(n)$ $\LCE$ queries on $w$ in $O(f(n))$ time for some function $f(n)$; then we can find all runs of $w$ in $O(n + f(n))$ time.\label{LCEtoRuns}
\end{lemma}

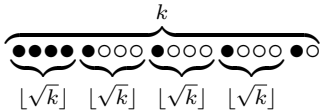
\begin{figure}[htb]
\centering
{\large
\begin{picture}(100,6)
\gasset{Nframe=n,Nw=5,Nh=3,AHnb=0,linewidth=0.25}
\put(39.5,2){\makebox(0,0)[cb]{$\overbrace{\phantom{%
{\bullet}{\bullet}{\bullet}{\bullet} {\bullet}{\circ}{\circ}{\circ} {\bullet}{\circ}{\circ}{\circ} {\bullet}{\circ}{\circ}{\circ} {\bullet}{\circ}{\circ}.}}^{k}$}}
\put(40,   -5){\makebox(0,0)[cb]{$\underbrace{{\bullet}{\bullet}{\bullet}{\bullet}}_{\lfloor\sqrt{k}\rfloor}%
\underbrace{{\bullet}{\circ}{\circ}{\circ}}_{\lfloor\sqrt{k}\rfloor}\underbrace{{\bullet}{\circ}{\circ}{\circ}}_{\lfloor\sqrt{k}\rfloor}%
\underbrace{{\bullet}{\circ}{\circ}{\circ}}_{\lfloor\sqrt{k}\rfloor}{\bullet}{\circ}$}}
\end{picture}
}
\caption{Simple difference cover of $[0..k)$ with $k = 18$.}\label{fig:simpleDC}
\end{figure}

In what follows we describe an algorithm that computes $O(n)$ $\LCE$ queries in $O(n\log^{\frac{2}3} n)$ time and thus prove Theorem using Lemma~\ref{LCEtoRuns}. The key notion in our construction is a \emph{difference cover}. Let $k\in \mathbb{N}$. A set $D \subset [0..k)$ is called a difference cover of $[0..k)$ if for any $x \in [0..k)$, there exist $y,z \in D$ such that $y - z \equiv x\pmod{k}$. Clearly $|D| \ge \sqrt{k}$. Conversely, for any $k \in \mathbb{N}$, there is a difference cover of $[0..k)$ with $O(\sqrt{k})$ elements: for example, the difference cover $[0..\lfloor\sqrt{k}\rfloor] \cup \{2\lfloor\sqrt{k}\rfloor, 3\lfloor\sqrt{k}\rfloor, \ldots\}$, which is depicted in Fig.~\ref{fig:simpleDC}. For further discussions and estimations of minimal difference covers, see \cite{ColbournLing,MereghettiPalano,Singer}.
\begin{example}
The set $D = \{1,2,4\}$ is a difference cover of~$[0..5)$.

$$
\begin{array}{c|c|c|c|c|c}
x & 0 & 1 & 2 & 3 & 4\\
\hline
y,z & 1,1 & 2,1 & 1,4 & 4,1 & 1,2
\end{array}
$$
\end{example}

Our algorithm utilizes the following interesting property of difference covers.
\begin{lemma}[see \cite{BurkhardtKarkkainen}]
Let $D$ be a difference cover of $[0..k)$. For any integers $i,j$, there exists $d \in [0..k)$ such that $(i + d) \bmod k \in D$ and $(j + d) \bmod k \in D$.\label{DiffCoverProperty}
\end{lemma}

\section{Longest Common Extensions}

At the beginning, our algorithm fixes an integer $\tau$ (the precise value of $\tau$ is given below). Let $D$ be a difference cover of $[0..\tau^2)$ of size $O(\tau)$. Denote $M = \{i \in [1..n] \colon (i \bmod \tau^2) \in D\}$. Obviously, we have $|M| = O(\frac{n}{\tau})$. Our algorithm builds in $O(\frac{n}{\tau}(\tau^2 + \log n)) = O(\frac{n}{\tau}\log n + n\tau)$ time a data structure that can calculate $\LCE(x, y)$ in constant time for any $x,y \in M$. To compute $\LCE(x, y)$ for arbitrary $x, y \in [1..n]$, we simply compare $w[x..n]$ and $w[y..n]$ from left to right until we reach positions $x+d$ and $y+d$ such that $x+d \in M$ and $y+d \in M$, and then we obtain $\LCE(x, y) = d + \LCE(x + d, y + d)$ in constant time. By Lemma~\ref{DiffCoverProperty}, we have $d < \tau^2$ and therefore, the value $\LCE(x, y)$ can be computed in $O(\tau^2)$ time. Thus, our algorithm can execute any sequence of $O(n)$ $\LCE$ queries in $O(\frac{n}{\tau}\log n + n\tau^2)$ time. Putting $\tau = \lceil\log^{\frac{1}3} n\rceil$, we obtain $O(\frac{n}{\tau}\log n + n\tau^2) = O(n\log^{\frac{2}{3}} n)$. Now it suffices to describe the data structure answering the $\LCE$ queries on the positions from $M$.

Let $i_1, i_2, \ldots, i_m$ be the sequence of all positions from $M$ in the increasing lexicographical order of the corresponding suffixes $w[i_1..n], w[i_2..n], \ldots, w[i_m..n]$. Our algorithm builds a \emph{longest common prefix array} $\lcp[1..m{-}1]$ such that $\lcp[j] = \LCE(i_j, i_{j+1})$ for $j \in [1..m)$ and a \emph{sparse suffix array} $\sa[1..n]$ such that $i_{\sa[x]} = x$ for $x \in M$ and $\sa[x] = 0$ for $x\notin M$. Obviously $\LCE(i_j, i_k) = \min\{\lcp[j], \lcp[j{+}1], \ldots, \lcp[k{-}1]\}$ for $j < k$. Based on this observation, we equip the $\lcp$ array with the \emph{range minimum query (RMQ)} structure~\cite{FischerHeun} that allows to compute $\min\{\lcp[j], \lcp[j{+}1], \ldots, \lcp[k{-}1]\}$ for any $j < k$ in $O(1)$ time. Now, to answer $\LCE(x, y)$ for $x, y \in M$, we first obtain $j = \sa[x]$ and $k = \sa[y]$ and then answer $\LCE(i_j, i_k)$ using the RMQ structure on the $\lcp$ array. Since the RMQ structure can be built in $O(n)$ time~\cite{FischerHeun}, it remains to describe how to construct $\lcp$ and $\sa$.

In general our construction is similar to that of~\cite{KosolobovLempelZiv}. We use the fact that the set $M$ has ``period'' $\tau^2$, i.e., for any $x\in M$, we have $x + \tau^2 \in M$ provided $x + \tau^2 \le n$. For simplicity, assume that $w[n]$ is a special letter that is smaller than any other letter in $w$. Our algorithm iteratively inserts the suffixes $\{w[x..n] \colon x \in M\}$ in the arrays $\lcp$ and $\sa$ from right to left. Suppose, for some $k \in M$, we have already inserted in $\lcp$ and $\sa$ the suffixes $w[x..n]$ for all $x \in M \cap (k..n]$. More precisely, denote by $i'_1, i'_2, \ldots, i'_{m'}$ the sequence of all positions $M\cap (k..n]$ in the increasing lexicographical order of the corresponding suffixes $w[i'_1..n], w[i'_2..n], \ldots, w[i'_{m'}..n]$; we suppose that $\lcp[j] = \LCE(i'_j, i'_{j+1})$ for $j \in [1..m')$, $i'_{\sa[x]} = x$ for $x \in M\cap(k..n]$, and $\sa[x] = 0$ for $x\notin M\cap(k..n]$. We are to insert the suffix $w[k..n]$ in $\lcp$ and $\sa$. In order to perform the insertions efficiently, during the construction, the arrays $\lcp$ and $\sa$ are represented by balanced search trees with some auxiliary structures as described below.

\paragraph{1. Balanced search tree for $\lcp$} The $\lcp$ array is represented by an augmented balanced search tree so that any RMQ query and modification on $\lcp$ take $O(\log n)$ amortized time.

\paragraph{2. List $L$} We store all positions $M\cap (k..n]$ on a linked list $L$ in the lexicographical order of the corresponding suffixes. We maintain on this list the order maintenance data structure of~\cite{BenderColeDemaineFarachColtonZito} that allows to determine whether a given node of $L$ precedes another node of $L$ in constant time. The insertion of a new node in $L$ takes amortized constant time. To provide constant time access to the nodes of $L$, we maintain an array $\nodes[1..n]$ such that $\nodes[x]$ is the node of $L$ corresponding to position $x$ if $x \in M\cap (k..n]$, and $\nodes[x] = \nil$ otherwise.

\paragraph{3. Balanced search tree for $\sa$} It is straightforward that, for any $x \in (k..n]$, $\sa[x]$ is equal to one plus the number of nodes of $L$ preceding $\nodes[x]$. So, we store all nodes of $L$ in an augmented balanced search tree allowing to calculate the number of nodes preceding $\nodes[x]$ in $O(\log n)$ time (since the comparison of two nodes takes $O(1)$ time). This tree together with the list $L$ and the array $\nodes$ allows to compute $\sa[x]$ in $O(\log n)$ time.

\paragraph{4. Trie $S$} We maintain a compacted trie $S$ that contains the strings $w[x..x{+}\tau^2]$ for all $x \in M \cap (k..n]$ (we assume $w[j] = w[n]$ for all $j > n$ and thus $w[x..x{+}\tau^2]$ is always well defined). We maintain on $S$ the data structure of~\cite{FranceschiniGrossi} supporting insertions in $O(\tau^2 + \log n)$ amortized time. Let $a$ be the leaf of $S$ corresponding to a string $w[x..x{+}\tau^2]$. We augment $a$ with a balanced search tree $B_a$ that contains nodes $\nodes[y]$ for all positions $y \in M\cap (k..n]$ such that $w[y{-}\tau^2..y] = w[x..x{+}\tau^2]$ (see Figure~\ref{fig:treeS}). We use $B_a$ to compute in $O(\log n)$ time the immediate predecessor and successor of any given node $\nodes[z]$, where $z \in M \cap (k..n]$, in the set of nodes stored in $B_a$. It is easy to see that $S$ together with the associated search trees occupies $O(\frac{n}{\tau})$ space in total.

\begin{example}
Let $\tau^2 = 4$. The set $D = \{0,1,3\}$ is a difference cover of $[0..\tau^2)$. Consider the string $w = \underline{a}b\underline{c}\underline{a}\underline{b}c\underline{a}\underline{b}\underline{a}b\underline{c}
\underline{a}\underline{b}b\underline{\$}$; the underlined positions are from $M = \{i \in [1..n] \colon (i\bmod \tau^2) \in D\}$. Figure~\ref{fig:treeS} depicts the compacted trie $S$; each leaf of $S$ is augmented with a balanced search tree of certain positions from $M\cap (k..n]$ (we use positions rather than nodes in this example). Consider the leaf of $S$ corresponding to the string $abcab$. The string $abcab$ occurs at positions $4, 9, 1$ in $w$. Hence, the balanced search tree $B_4$ must contain three positions: $4{+}\tau^2 = 8, 9{+}\tau^2 = 13, 1{+}\tau^2 = 5$. Note that the positions are stored in the lexicographical order of the corresponding suffixes $w[8..n], w[13..n], w[5..n]$.
\begin{figure}[htb]
\includegraphics[scale=0.35]{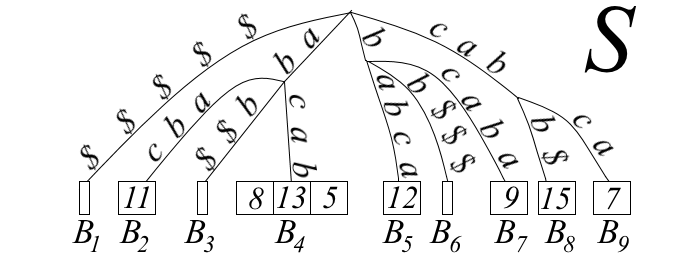}
\caption{The balanced search trees $B_1, B_2, \ldots, B_9$ are augmented with some positions from $M$.}
\label{fig:treeS}
\end{figure}
\end{example}

\paragraph{The construction of $\lcp$ and $\sa$} To insert $w[k..n]$ in $\lcp$ and $\sa$, we first insert $w[k..k{+}\tau^2]$ in $S$ in $O(\tau^2 + \log n)$ time. If $S$ did not contain the string $w[k..k{+}\tau^2]$ before, then, using auxiliary structures on $S$, we easily find in $O(1)$ time the position in $\lcp$ where the suffix $w[k..n]$ should be inserted; in the same way we obtain the $\LCE$ value between $w[k..n]$ and its immediate predecessor and successor in $S$. Then, we modify the balanced search tree representing $\lcp$, insert a new node corresponding to $w[k..n]$ in $L$, insert this node in the balanced search tree supporting $\sa$, and, finally, add a new empty tree $B_a$ to the newly created leaf $a$ of $S$. All these modifications take $O(\log n)$ amortized time.

Now suppose $S$ contains $w[k..k{+}\tau^2]$. Denote by $a$ the leaf of $S$ corresponding to $w[k..k{+}\tau^2]$. In $O(\log n)$ time we obtain the immediate predecessor and successor of the node $\nodes[k{+}\tau^2]$ (recall that $k{+}\tau^2 \in M$) in the search tree $B_a$; denote these nodes by $\nodes[x]$ and $\nodes[y]$, respectively. (We assume that the predecessor and successor both are defined; the case when one of them is undefined is analogous). Note that $\nodes[x]$ is the immediate predecessor only in the set of all nodes contained in $B_a$ but it may not be the immediate predecessor in the whole list $L$; the situation with $\nodes[y]$ is similar. Then we insert $\nodes[k{+}\tau^2]$ between $\nodes[x]$ and $\nodes[y]$ in $B_a$. Since $w[x{-}\tau^2..x] = w[y{-}\tau^2..y] = w[k..k{+}\tau^2]$, it is straightforward that the suffixes $w[x{-}\tau^2..n]$ and $w[y{-}\tau^2..n]$ are, respectively, the immediate predecessor and successor of the suffix $w[k..n]$ in the set of all suffixes $\{w[x..n] \colon x\in M\cap (k..n]\}$. Hence, we insert a new node $\nodes[k]$ in $L$ between the nodes $\nodes[x{-}\tau^2]$ and $\nodes[y{-}\tau^2]$ (these nodes are certainly adjacent).

It is easy to see that $\LCE(k, x{-}\tau^2) = \tau^2 + \LCE(k{+}\tau^2, x)$ and $\LCE(k, y{-}\tau^2) = \tau^2 + \LCE(k{+}\tau^2, y)$. The values $\LCE(k{+}\tau^2, x) = \LCE(i'_{\sa[k{+}\tau^2]}, i'_{\sa[x]})$ and $\LCE(k{+}\tau^2, y) = \LCE(i'_{\sa[k{+}\tau^2]}, i'_{\sa[y]})$ can be computed in $O(\log n)$ time using the balanced search trees supporting access on $\sa$ and RMQ queries on $\lcp$. All subsequent changes of other structures are the same as in the previous case and require $O(\log n)$ amortized time.

Finally, once the last suffix is inserted, we construct in an obvious way the plain arrays $\lcp$ and $\sa$ in $O(n)$ time.

\paragraph{Time and space} The insertion of a new suffix in the arrays $\lcp$ and $\sa$ takes $O(\tau^2 + \log n)$ amortized time. Thus, the construction of $\lcp$ and $\sa$ consumes overall $O(\frac{n}{\tau}(\tau^2 + \log n))$ time as required. The whole data structure occupies $O(n)$ space.

\section{Conclusion}

It seems that further improvements in the considered problem may be achieved by more efficient longest common extension data structures on a general ordered alphabet. One even might conjecture that there is a data structure that can execute any sequence of $k$ $\LCE$ queries on a string of length $n$ over a general ordered alphabet in $O(k + n)$ time. However, we do not yet have a theoretical evidence for such strong results.

Another interesting direction is a generalization of our result for the case of online algorithms (e.g., see~\cite{HongChen} and~\cite{Kosolobov2}).

\subparagraph*{Acknowledgements}
The author would like to thank Gregory Kucherov for inviting in Universit{\'e} Paris-Est, where the present result was obtained, and the anonymous referee who simplified the proof and highly improved the quality of the paper.

\section*{References}

\bibliography{faster_runs}

\end{document}